\documentclass[1p,review]{elsarticle}
\pdfoutput=1 
 						
\usepackage[final]{changes}  

\usepackage{subcaption}
\usepackage{wrapfig}
\usepackage{xcolor}
\usepackage{lineno}
\usepackage{amsmath}
\usepackage[tableposition=top]{caption}
\usepackage{siunitx}
\DeclareSIUnit\samplepers{Sps}
\DeclareSIUnit\photoelectron{PE}
\DeclareSIUnit\bit{b}
\DeclareSIUnit\ton{t}
\DeclareSIUnit\DACc{DAC~counts}

\usepackage[hidelinks, colorlinks=true]{hyperref}

\title{\boldmath Mass testing of the JUNO experiment\\ 20-inch PMTs readout electronics}

\author[a]{Alberto~Coppi}
\author[a,b]{Beatrice~Jelmini\corref{cor1}}
\cortext[cor1]{Corresponding author}
\author[b]{Marco~Bellato}
\author[b]{Antonio~Bergnoli}
\author[a,b]{Matteo~Bolognesi}
\author[a,b]{Riccardo~Brugnera}
\author[a]{Vanessa~Cerrone}
\author[c]{Chao~Chen}
\author[d]{Barbara~Clerbaux}
\author[b]{Daniele~Corti}
\author[b]{Flavio~dal~Corso}
\author[e]{Jianmeng~Dong}
% ...
\cortext[cor2]{Now at INFN Laboratori Nazionali del Sud, Italy}
\author[a,b]{Chiara~Sirignano}
\author[l]{Monica~Sisti}
\author[b]{Luca~Stanco}
\author[i,j]{Virginia~Strati}
\author[h,g]{Francesco~Tortorici}
\author[h,g]{Cristina~Tuv\'e}
\author[m]{Carlo~Venettacci}
\author[g]{Giuseppe~Verde}
\author[p]{Lucia~Votano}
\address[g]{INFN Sezione di Catania, Catania, Italy}
\address[k]{INFN Sezione di Milano e Universit\`a di Milano, Dipartimento di Fisica, Milano, Italy}
\address[l]{INFN Sezione di Milano Bicocca, e Universit\`a di Milano Bicocca, Dipartimento di Fisica, Milano, Italy}
\address[m]{INFN Sezione di Roma Tre e Universit\`a di Roma Tre, Dipartimento di Matematica e Fisica, Roma, Italy}
\address[n]{INFN, Sezione di Milano Bicocca e Politecnico di Milano, Dipartimento di Energetica, Milano, Italy}
\address[o]{INFN Sezione di Perugia e Universit\`a di Perugia, Dipartimento di Chimica, Biologia e Biotecnologie, Perugia, Italy}
\address[p]{Laboratori Nazionali dell'INFN di Frascati, Italy}
\address[h]{Universit\`a di Catania, Dipartimento di Fisica e Astronomia, Catania, Italy}
\address[i]{INFN Sezione di Ferrara, Ferrara, Italy}
\address[j]{Universit\`a degli Studi di Ferrara, Dipartimento di Fisica e Scienze della Terra, Italy}

\begin{document}

\begin{abstract}
The Jiangmen Underground Neutrino Observatory (JUNO) is a multi-purpose, large size, liquid scintillator experiment under construction in China.
JUNO will perform leading measurements detecting neutrinos from different sources (reactor, terrestrial and astrophysical neutrinos) covering a wide energy range (from \SI{200}{\keV} to several \si{\GeV}).
This paper focuses on the design and development of a test protocol for the 20-inch PMT underwater readout electronics, performed in parallel to the mass production line. 
In a time period of about ten months, a total number of 6950 electronic boards were tested with an acceptance yield of \SI{99.1}{\percent}.
\end{abstract}


\maketitle

\section{Introduction}
\label{sec:intro}

The Jiangmen Underground Neutrino Observatory~\cite{bib:juno:phys-det} (JUNO) is a \SI{20}{kton} neutrino medium baseline experiment under construction in southern China.
The JUNO experiment has been proposed~\cite{bib:juno:yb} with the main goals of determining the neutrino mass ordering with a significance of $3\,\sigma$ within the first six years of data taking and measuring the oscillation parameters, $\Delta m^2_{21}$, $\Delta m^2_{31}$, and $\sin^2 \theta_{12}$, with sub-percent precision~\cite{bib:precision}.
To achieve these goals, JUNO is located about \SI{53}{km} away from two nuclear power plants and will detect electron antineutrinos produced by the beta decays of fission products inside the nuclear cores.
JUNO will also be able to address many other topics in particle and astroparticle physics, by detecting neutrinos from natural sources: solar neutrinos, atmospheric neutrinos, geo-neutrinos, neutrinos from core-collapse supernovae, and from the diffuse supernovae neutrino background.
An updated overview of the JUNO physics reach can be found here\,\cite{bib:juno:phys-det}. 

The JUNO Central Detector (CD) consists of \SI{20}{kton} of liquid scintillator contained in a spherical acrylic vessel with a \SI{35.4}{\m} diameter, supported by a stainless Steel Truss. 
A double system of \num{17612} 20-inch large-PMTs (LPMTs)~\cite{bib:juno_pmt,bib:pmt_testing} and \num{25600} 3-inch small-PMTs (SPMTs)~\cite{CAO2021165347} is employed to detect the scintillation and Cherenkov light produced by neutrino interactions with the liquid scintillator. 
The liquid scintillator target is surrounded by a 35-kt pure Water Pool, which is instrumented with 2400 20-inch LPMTs; the Water Pool shields the inner part of the detector from environmental radioactivity, and is part of the muon Veto system, together with the Top Tracker on top of the whole structure. 

The JUNO LPMT underwater readout electronic system is responsible to sample and to process the LPMT output current~\cite{bib:elec:bx, bib:electronics}. 
Spotting hardware failures and evaluating the performance of the underwater readout electronics before the actual installation is of paramount importance, because it will be impossible to repair or to change an electronics module after its deployment.
Furthermore, the required loss rate of the electronics channels is
less than 0.5\% in 6 years~\cite{bib:juno:phys-det}.
To this end, we designed and developed a dedicated test protocol~\cite{thesis:Coppi} to be carried out during the mass production, held in a dedicated facility in Kunshan, China.

The rest of the paper is organized as follows: in Section~\ref{sec:electronics} we describe the JUNO LPMT readout electronics; in Section~\ref{sec:kunshan} we discuss the mass production and the mass testing setup at the dedicated facility in Kunshan; in Section~\ref{sec:test_protocol} a detailed description of the developed test protocol is presented; conclusions are drawn in Section~\ref{sec:conclusion}.

\section{JUNO LPMT readout electronics}
\label{sec:electronics}

\begin{figure}[htbp]
\centering
  \includegraphics[width=1\columnwidth]{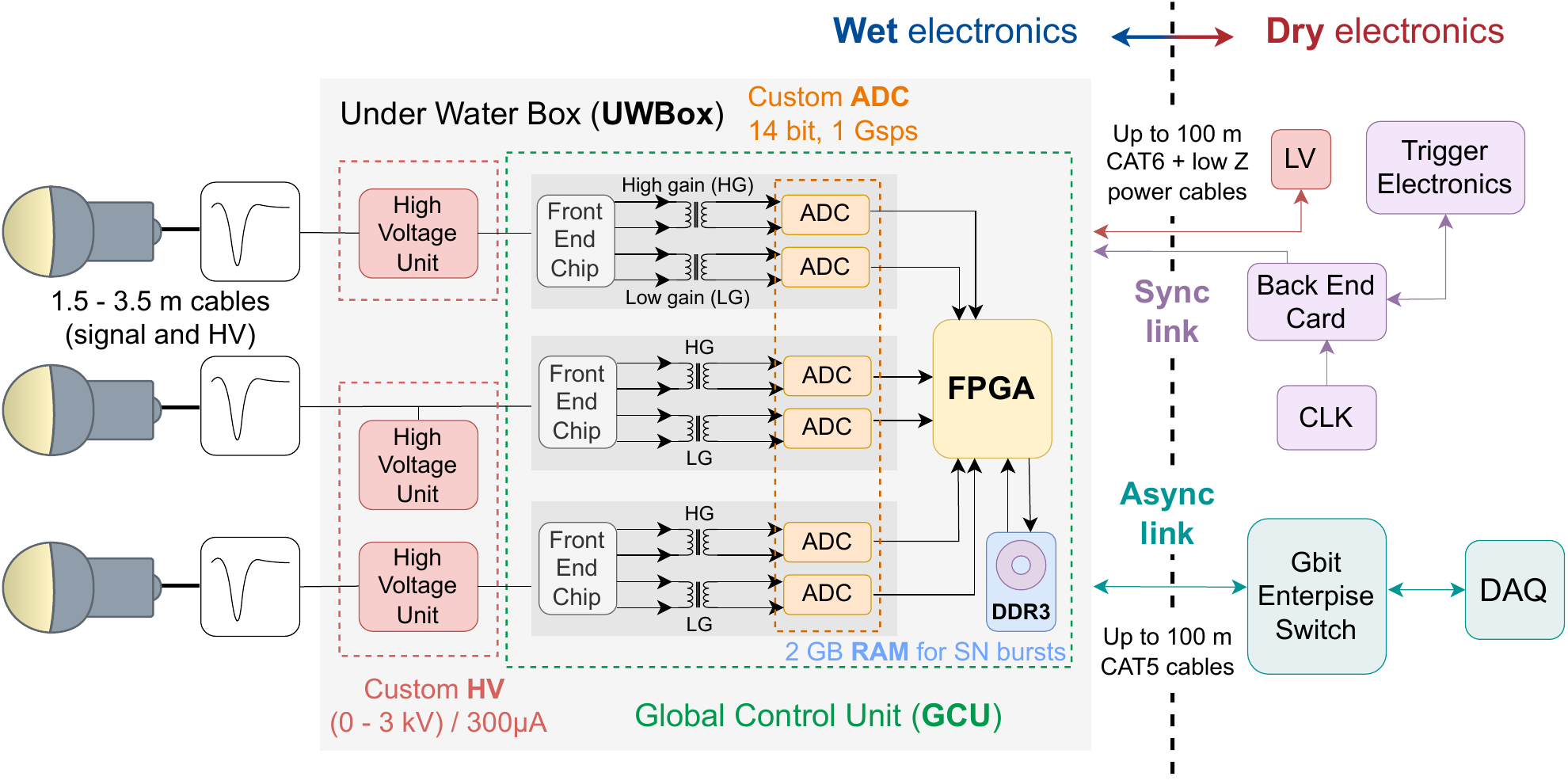}
  \caption{\label{fig:1f3:scheme}JUNO LPMT readout
           electronics scheme. A description of the different parts is given in the text.}
\end{figure}

A scheme of the JUNO LPMT electronics is given in Figure~\ref{fig:1f3:scheme}~\cite{bib:electronics}; the design is an optimization of previous developments~\cite{bib:elec:bx}.
The full electronics chain is composed of two parts: the \emph{front-end} (FE), or \emph{wet}, electronics~\cite{thesis:Marini} located very close to the LPMT output, inside the JUNO Water Pool; and the \emph{dry} electronics, installed in the electronics rooms of the JUNO underground laboratories, which consists of the \emph{back-end} (BE), or trigger, electronics and the data acquisition (DAQ) system.

The FE electronics will be installed underwater on the JUNO Steel Truss structure, inside a stainless steel, water-tight box, the so-called Under Water Box (UWBox).
In total, the JUNO detector is instrumented with 6681 UWBoxes, 5878 for the CD and 803 for the Water Pool as part of the JUNO Veto system.
Three LPMT output signals are fed to one UWBox which contains
\begin{itemize}
\item three High Voltage Units (HVU): programmable modules which provide the bias voltage to the LPMT voltage divider. 
Each HVU independently powers one LPMT. 
The HVUs are mounted on a custom Printed Circuit Board (PCB), the splitter board, that provides mechanical stability to the modules, and decouples the PMT signal current from the high voltage.
\item a Global Control Unit (GCU): a motherboard incorporating the front-end and readout electronics components. 
The three LPMT signals reaching the GCU are processed though independent readout chains.
\end{itemize}

The LPMT analog signal reaching the GCU is processed by a custom Front-End Chip (FEC), which duplicates the input signal and inject it in two parallel streams with different gains, referred to as \emph{high-gain stream} and \emph{low-gain stream} (see Figure~\ref{fig:1f3:scheme}).
The signal from each stream is further converted to a digital waveform by a 14-bit, 1\,GS/s, custom Flash Analog-to-Digital Converter (FADC). 

The usage of two FADCs per readout channel has been driven by the design requirement of providing a wide dynamic range in terms of reconstructed photo-electrons (PE): from \SI{1}{\photoelectron} to \SI{100}{\photoelectron} (high-gain stream) with a resolution of \SI{0.1}{\photoelectron} or better, and from \SI{100}{\photoelectron} to \SI{1000}{\photoelectron} (low-gain stream) with a resolution of \SI{1}{\photoelectron} or better~\cite{bib:juno:yb, Liu:2022nhe}, with a nominal LPMT gain of \num{e7}.

A Xilinx Kintex-7 FPGA (XC7K325T) is the core of the GCU and allows to further process the digital signal (local trigger generation, charge reconstruction and timestamp tagging) and temporarily store it in a local memory buffer before sending it to the data acquisition (DAQ).
Besides the local memory available in the readout-board FPGA, a \SI{2}{GBytes} DDR3 memory is available and used to provide a larger memory buffer in the exceptional case of a sudden increase of the input rate, which overruns the current data transfer bandwidth between the FE electronics and the DAQ~\cite{thesis:Triozzi}.

The BE electronics is composed of the following active elements:
\begin{itemize}
\item the Back End Card (BEC) with the Trigger and Time Interface Mezzanine
      (TTIM)
\item the Reorganize and Multiplex Units (RMU) and the Central Trigger
      Unit (CTU), which are part of the Trigger Electronics
      (see Figure~\ref{fig:1f3:scheme}).
\end{itemize}

The LPMTs are connected to the UWBox electronics with a \SI{50}{\ohm}, coaxial cable, with a length ranging between \SI{1.5}{m} and \SI{3.5}{m}. 
The electronics inside the UWBox has two independent connections to the dry electronics: a so-called \emph{synchronous link} (S-link) for the connection to BE electronics, which provides the clock and synchronization to the boards and handles the trigger primitives; and an \emph{asynchronous link} (A-link) which is fully dedicated to the DAQ and slow-control, or Detector Control System (DCS). 
These connections are realized using commercially available CAT-5 and CAT-6 Ethernet cables for the A-link and S-link, respectively; the length of the cables ranges between \SI{30}{m} and \SI{100}{m}.
An additional, low-resistance, power cable will be used to bring power to the electronics inside the UWBox.

The LPMT electronics can run with a centralized \emph{global trigger} mode, where the information from the single fired PMTs is collected and processed in the CTU. 
The latter validates the trigger based on a simple PMT multiplicity condition or a more refined topological distribution of the fired LPMTs in JUNO~\cite{Gong:2015yvq}.
Upon a trigger request, validated waveforms are sent to the DAQ event builder through the A-link. 
The IPBus Core protocol~\cite{bib:ipbus} is used for data transfer~\cite{bib:Triozzi}, slow control monitoring, and electronics configurations.

An alternative scheme is possible where all readout boards send their locally triggered waveforms to the DAQ, independently of each other, without passing through the BE trigger electronics and the S-link. 
With this approach, all the digitized waveforms, including those generated by dark noise photo-electrons, would be sent to the DAQ. 

The BE electronics was only used in the tests presented in this paper to provide the UWBox with the clock needed to operate properly, while it was not used to handle trigger information; hence, the boards were operated in the locally triggered setup explained above.
We chose this configuration because we used the internal test pulse generator, described in the next section, to generate the input signals, but the internal generator could only be activated through the A-link, thus generating non-synchronized waveforms.
So, for the purpose of the electronics mass testing, it was easier to just read every waveform without relying on any triggering logic.

\subsection{Internal test pulse generator}\label{subsec:testpulser}

\begin{figure}[hbtp] 
  \centering
  \includegraphics[width=1\columnwidth]{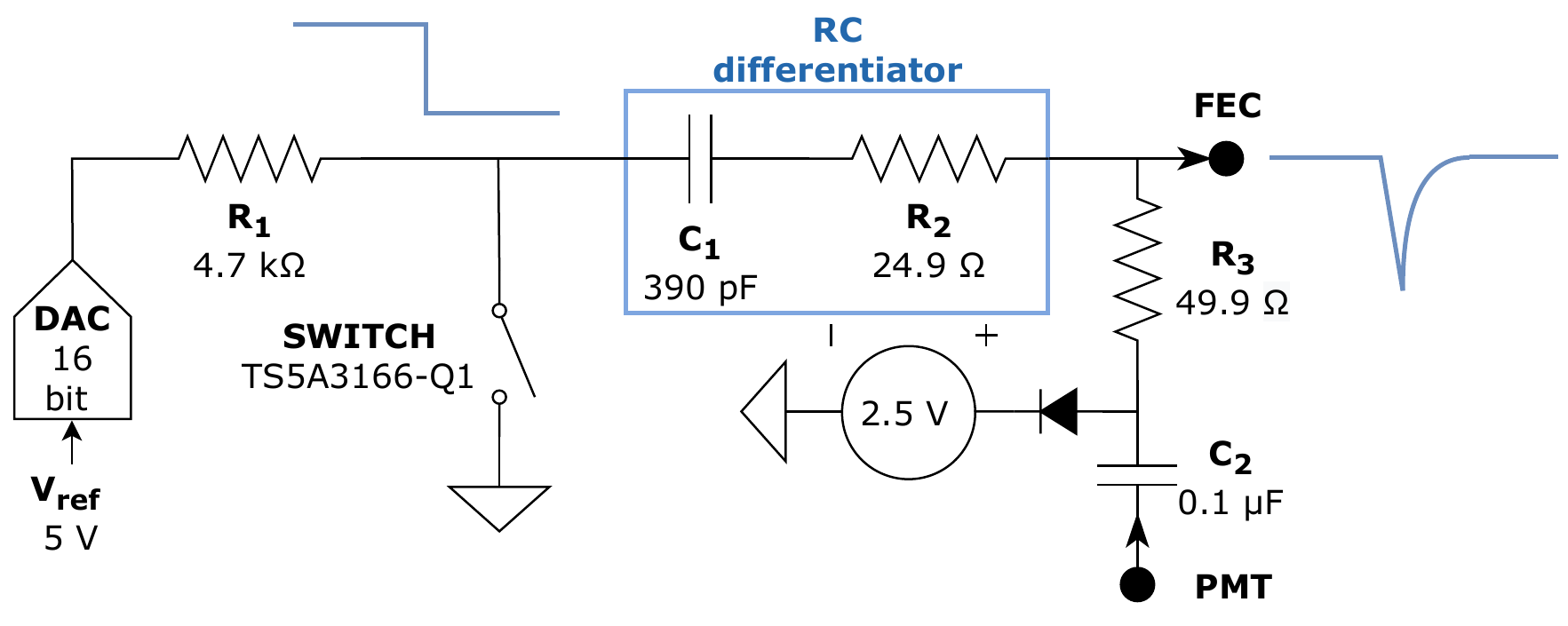}
  \caption{Scheme of the internal test pulse generator. Each channel is equipped with one internal generator circuit, which is connected directly to the Front-End Chip (FEC). The main components of the circuit are a 16-bit digital-to-analog converter (DAC), a switch, and a RC circuit with $C_1 = \SI{390}{\pico\farad}$ and $R_2 = \SI{24.9}{\ohm}$. The connection from the PMT to the FEC is also shown; arrows are used to indicate the direction of the signals.}
\label{fig:scheme_testpulse}
\end{figure}

Each GCU is equipped with three test pulse generator circuits, one per channel; a scheme of the circuit is presented in Figure~\ref{fig:scheme_testpulse}.
The main components of the circuit are a 16-bit digital-to-analog converter (DAC), a switch, and a RC circuit acting as a differentiator, or high-pass filter, with $C_1 = \SI{390}{\pico\farad}$ and $R_2 = \SI{24.9}{\ohm}$, with a $\SI{5}{\percent}$ and $\SI{1}{\percent}$ tolerances, respectively; the values of $C_1$ and $R_2$ were chosen to produce a signal mimicking a PMT signal.

The amplitude of the generated pulse can be adjusted via IPbus protocol~\cite{bib:ipbus} by changing the input digital amplitude of the DAC ($A_{\text{DAC}}$), which uses a reference voltage of \SI{5}{\volt} to convert the digital value to a voltage value.
The pulse is generated by closing the switch and connecting the node between the DAC and the differentiator to ground, generating a step voltage, as shown in Figure~\ref{fig:scheme_testpulse}. 
The step function goes through a differentiator, or high-pass filter, generating a PMT-like pulse which is injected directly into the FEC of the channel.
The switch is also controlled via the IPbus protocol: to generate one pulse, we need to close and then open again the switch, hence two IPbus commands are needed; in this way it is possible to control the frequency at which the switch is closed/open and the test pulses are generated.

The injected input charge, which is the area of each generated pulse, corresponds to the charge accumulated by the capacitor $C_1$ under a potential difference equal to the DAC output, evaluated as follows:
\begin{linenomath}
\begin{equation}
Q_{\text{in}} = A_{\text{DAC}} \cdot \frac{ \SI{5}{\volt} }{ 2^{16} } \cdot C_1,
\label{eqn:input_charge}
\end{equation}
\end{linenomath}
where $Q_{\text{in}}$ is in unit of \si{\pico\coulomb} if $C_1$ and $A_{\text{DAC}}$ are in units of \si{\pico\farad} and \si{\DACc}, respectively. 
The value $\SI{5}{\volt}  / 2^{16} \simeq \SI{76}{\micro\volt}$/DAC counts is the conversion factor from DAC counts to a tension in volts.

\section{Mass production and testing at the Kunshan site}\label{sec:kunshan}

A facility in Kunshan, China, was devoted to the mass production and testing of the 20-inch PMTs readout electronics.

\subsection{Production process}
\label{subsec:production}

\begin{figure}[hbtp] 
  \centering
		\subfloat[][\emph{Assembled UWBox before laser welding.}]
		   {\includegraphics[width=.5\columnwidth]{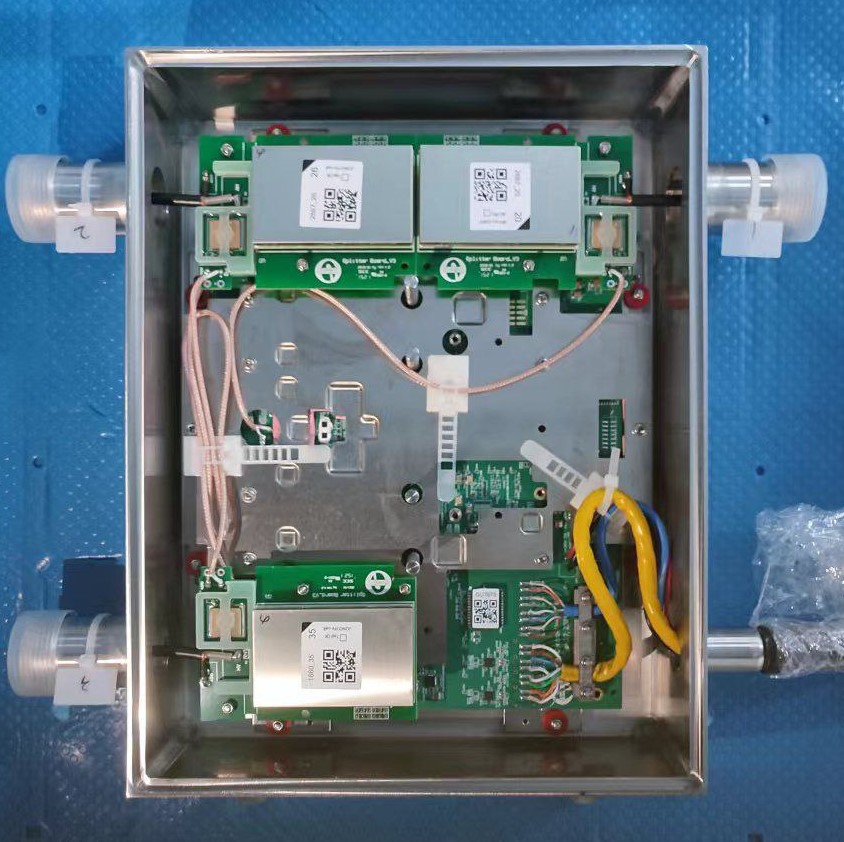}
		   \label{subfig:uwbox}} \quad
		\subfloat[][\emph{Shelf with UWBoxes in the testing room in Kunshan.}]
		   {\includegraphics[width=.65\columnwidth]{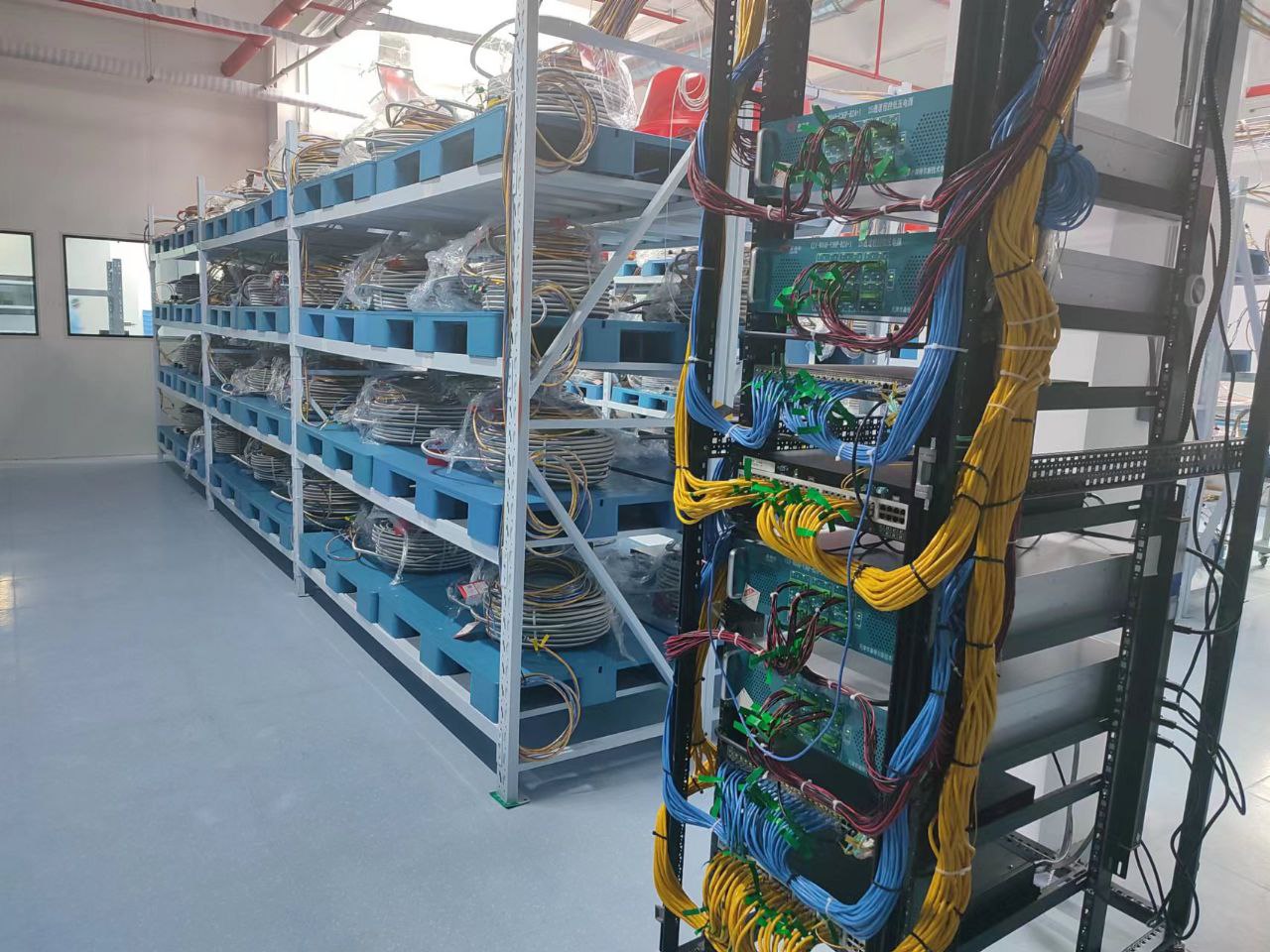}
		   \label{subfig:testing_room}} \\

  \caption{(a) Picture of an assembled Underwater Box before laser welding. The three HVUs are clearly visible, one near each of the connectors at which the LPMTs will be connected. The GCU board is located on the bottom. (b) A shelf full of UWBoxes in the testing room at the Kunshan facility. In the front of the picture, a rack with power supplies, switches, and back-end and trigger electronics is also visible.}
\label{fig:kunshan_pictures}
\end{figure}

During mass production, the first step was the welding of the stainless steel bellow to the UWBox, followed by a leakage test. 
After that, the threading of the cables for the S-link, A-link and the power line through the bellow was done.
Then, the GCU board and the three HV units were assembled inside the threaded UWBox and soldered.
Each box was then tested for at least five days, and if it passed the tests, it was finally laser welded, and, after a leakage test, was stored before being sent to the JUNO experimental site.
A picture of an assembled UWBox before laser welding is shown in Figure~\ref{subfig:uwbox}.

Before the beginning of the mass production, tests were performed on a small number of boxes to assess the possible damage and risks from the laser welding procedure; it was found that no damage is expected from this procedure.
Nonetheless, a quicker version of the tests was performed on each board after the laser welding.

\subsection{Testing of the GCUs}
\label{subsec:testing}

During the test, the assembled UWBoxes and the bellows were located on shelves in a dedicated testing room, as shown in Figure~\ref{subfig:testing_room}; in the front of the picture, a rack with power supplies, switches for the network connection, and the trigger electronics is also visible.
The room had places to locate a maximum number of 344 GCUs on nine shelves.
All the tests described in Section~\ref{sec:test_protocol} were performed in parallel on all the GCUs available in the testing room.

\begin{figure}[hbtp] 
  \centering
  \includegraphics[width=1.\columnwidth]{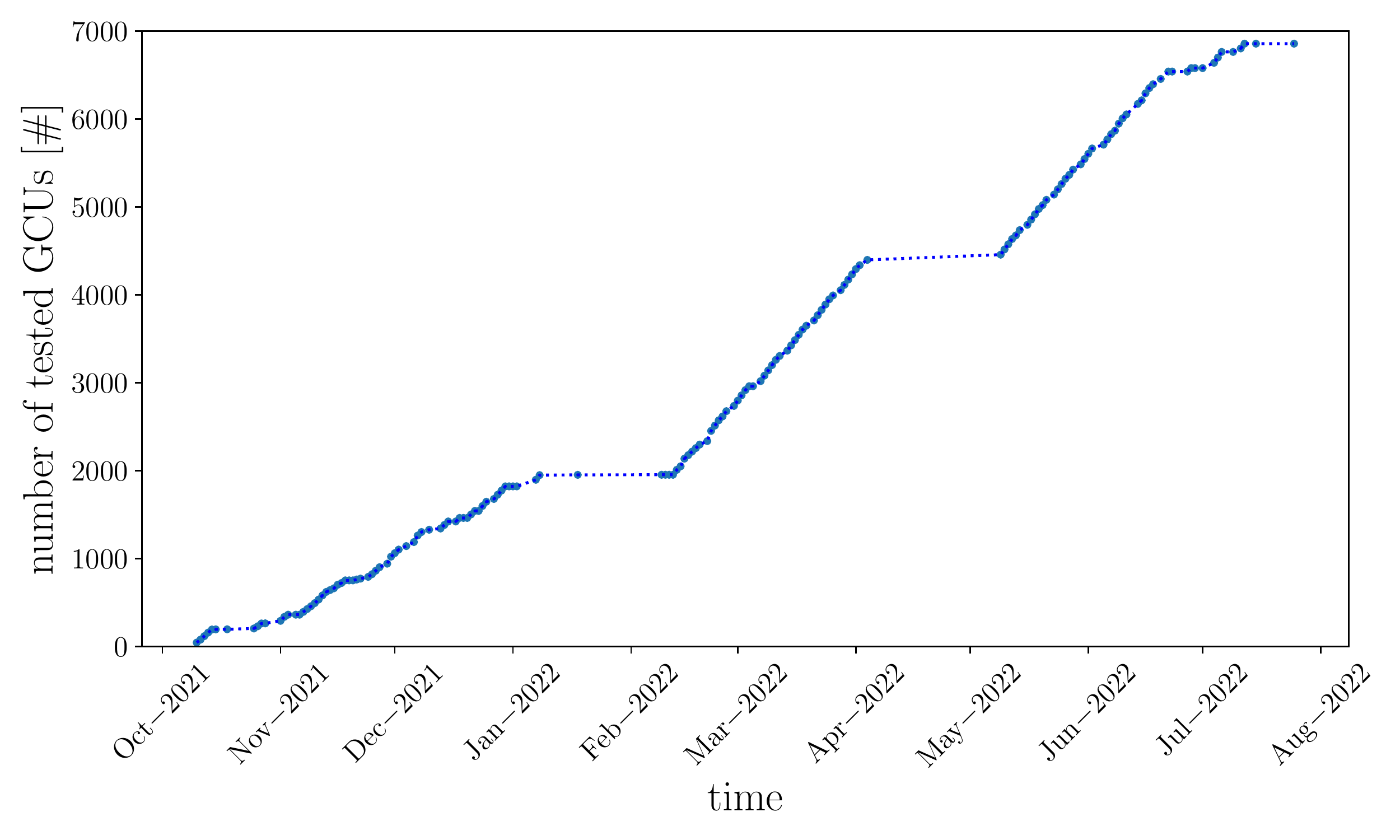}
  \caption{Cumulative number of tested GCUs as a function of time. The production and testing campaign started in October 2021 and ended in July 2022. A total of 6900 boards were tested. Two breaks in the production, the first due to Chinese New Year Holidays and the second due to a COVID-19 outbreak, are clearly visible.}
\label{fig:tested_gcus}
\end{figure}

The test procedure was automatized in order to minimize human errors during the shifts.
Shifts were organized exploiting time zone differences between China, where the boxes were located, and Europe, so that the European part of the collaboration could take part in the mass testing remotely, since it was not possible to travel to China due to COVID-19 restrictions.
During day time in China, local shifters were in charge of assembling between 40 and 60 new UWBoxes per day and replacing them in the testing room; at the end of the Chinese working day, an European shifter took over to perform the tests, in this way it was possible to have shifts covering all 24 hours each day.
Data analysis on the acquired data from the tests was performed on the following day, in order to provide a fast feedback on the tested boards.
The mass testing of all 6950 GCUs lasted for about 10 months from October 2021 to July 2022.
Figure~\ref{fig:tested_gcus} shows the cumulative number of tested boards as a function of time.

\subsection{Network and connection details at the Kunshan site}
\label{subsec:connections}

In the testing room, GCUs were connected to the BECs in batches of 40 in order to provide the clock to the tested boards through the synchronous link.
For the asynchronous link, 40 GCUs were connected to a level 1 (L1) switch through a \SI{1}{\giga\bit} link, for a total of nine L1 switches; L1 switches were then connected to a level 2 (L2) switch through a 4x\SI{10}{\giga\bit} link; the L2 switch was finally connected to the DAQ server via a 4x\SI{100}{\giga\bit} link. 
The DAQ server consisted of a Dell PowerEdge C6400, with a total of 24 cores and 48 threads, \SI{2.7}{\giga\hertz} processor and \SI{192}{GB} RAM.
A dedicated local network was used for the communication between the GCUs and the server.

\section{Test Protocol for the LPTM readout electronics}
\label{sec:test_protocol}

We designed and implemented the test protocol~\cite{thesis:Coppi} according to the following criteria: (1) it had to be controlled remotely and to be run in parallel to the production line; (2) it had to be easy to operate, in order to have non-expert shifters being able to join the testing campaign; (3) it had to provide the shifter with a fast and visual feedback of the performance of the tested components.

The test protocol was performed on each electronics card after all the components got assembled together, as described in Section~\ref{subsec:production}, and before the UWBox was finally sealed by means of laser welding and then sent to a storage warehouse near the JUNO site.
It is made of several steps: (1) a ping test (sec.~\ref{subsec:ping}), to check the connection of the board to the local network; (2) a linearity (sec.~\ref{subsec:linearity}) and a stability (sec.~\ref{subsec:stability}) tests investigating the properties of the digitized waveforms to validate the performance and the reliability of the whole readout chain; (3) a DCS test (sec.~\ref{subsec:slowcontrol}) to monitor the temperature and the status of the board.
Each test is presented in more details in the following subsections.

The tests of step (2) were performed separately on the high-gain and on the low-gain streams. 
Input signals were generated in both cases by the internal test pulse generator, but either the high-gain stream, or the low-gain stream was selected for the readout of the digitized waveform. 

\subsection{Properties of the digitized waveform}\label{subsec:waveform}

\begin{figure}[hbtp] 
  \centering
  \includegraphics[width=1.\columnwidth]{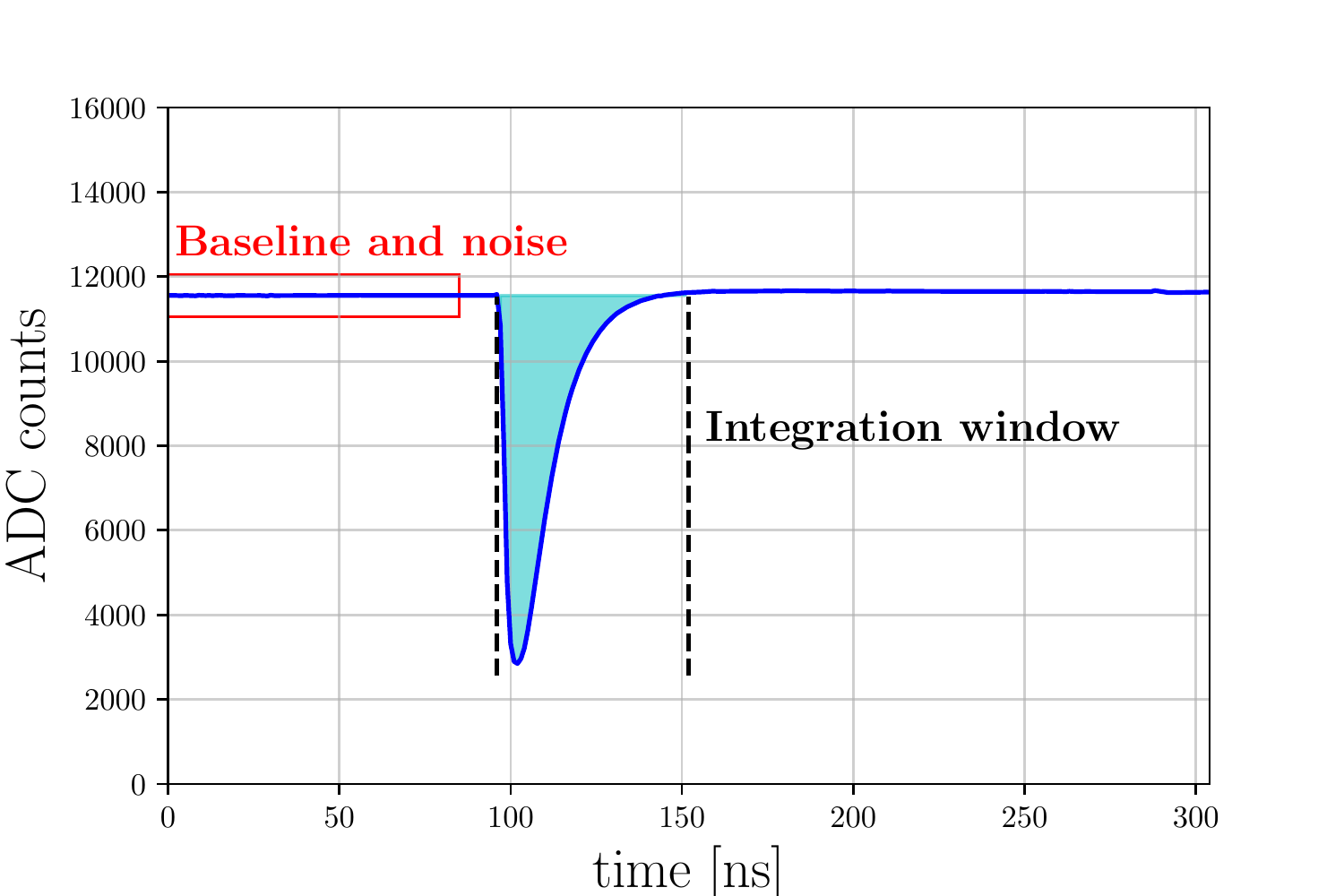}
  \caption{Example of a digitized waveform from GCU 3133 channel 0, generated with the internal test pulse generator, as described in Section~\ref{subsec:testpulser}, and obtained by selecting the high-gain stream. The first 85 samples are used to evaluate the values of the baseline and the noise. The limits of the charge integration window are shown as dashed black lines.}
\label{fig:waveform}
\end{figure}

Figure~\ref{fig:waveform} shows an example of a digitized waveform generated by the internal test pulse generator described in Section~\ref{subsec:testpulser}, where the high-gain stream was selected.
During the tests, the length of the readout window, and hence the length of the waveform, is fixed to 304 samples which correspond to \SI{304}{\nano\second}, given the FADC sampling frequency of \SI{1}{GS/s}.

For each digitized waveform, baseline and noise are evaluated.
The baseline, $B$, is defined as the average of the first 85 samples; the noise, $\sigma_{\text{baseline}}$, is defined as the standard deviation computed on the same samples.

Another property which is monitored during the test is the waveform integrated charge. 
The waveform integrated charge, $Q_{\text{out}}$, corresponds to the shadowed region in Figure~\ref{fig:waveform} and it is evaluated offline as in the following equation:
\begin{linenomath}
\begin{equation}
Q_{\text{out}} = \sum_i^{N_s} \frac{ \SI{75}{\micro\volt} \cdot  \left( B - N_i \right) \cdot \Delta t_s}{ R }, 
\label{eqn:output_charge}
\end{equation}
\end{linenomath}
where $N_s$ is the number of bins in the integration window, $N_i$ is the amplitude in ADC counts of the $i$-th bin, $B$ the baseline value as defined above, $\SI{75}{\micro\volt} $ is the voltage corresponding to 1 ADC count, $R = \SI{50}{\ohm}$ is the input impedance, and $\Delta t_s$ is the width of a single bin; in our case $\Delta t_s = \SI{1}{\nano\s}$.
The integration window, shown in Figure~\ref{fig:waveform}, starts \SI{5}{\nano\second} before the minimum, or peak, of the waveform, up to \SI{50}{\nano\second} after the minimum.

In eq.~\eqref{eqn:output_charge}, the conversion factor between ADC counts and voltage, $\SI{75}{\micro\volt} $/ADC count, is a characteristic of the FADCs which depends on the dynamic range and the number of bits, and it is the same for the high-gain and low-gain streams.
In this way, eq.~\eqref{eqn:output_charge} does not take into account the gain of the amplification step in the FEC, which in turn has to be determined through the linearity test of the test protocol, as explained in Section~\ref{subsec:linearity}.

\subsection{Configuration of the GCUs}\label{subsec:setup}

The following GCU parameters needed to be set through the slow control before each test: (1) the length of the readout window; (2) the value of the pre-trigger; (3) the value of the trigger threshold; (4) and the trigger mode.
For the mass production tests, we fixed the length of the readout window to \SI{304}{\nano\second} to optimize the total size of the acquired data. 
The pre-trigger is the time interval between the moment at which the signal exceeds the threshold and the beginning of the readout window, \emph{i.e.}, the region that precedes the pulse.

There are two possibilities for the trigger threshold: the threshold is either fixed to a given value in ADC counts, and is the same for all channels; or it is evaluated for each channel in terms of $\sigma_{\text{baseline}}$ from the baseline.
During the tests, the trigger threshold was fixed to a common value for all channels.
The trigger modes have already been described in Section~\ref{sec:electronics}; during the tests, the trigger mode was set to locally triggered approach, therefore each channel triggers independently from each other and the BE trigger electronics is not employed.

\subsection{Ping test}\label{subsec:ping}

The first step of the test protocol is the ping test, meant to check that all the GCUs are properly connected to the local network and responding.
A non-responding board would imply either that the cables are not properly plugged in, which is an easy issue to solve, or that the assembling procedure was not successful, thus requiring further investigation on the production side.

For this test, we used the default Linux \texttt{ping} command and sent 100 56-byte packets in \SI{1}{\s} from the DAQ server to each GCU, so that it was possible to test in a few seconds the connection to the local network of hundreds of boards; the IP addresses were automatically recovered by the input GCU ID number. 
Note that the \texttt{ping} command directly calculates the mean response time and its standard deviation, which were both stored, together with the fraction of lost packets.

\begin{figure}[hbtp] 
  \centering
  \includegraphics[width=1\columnwidth]{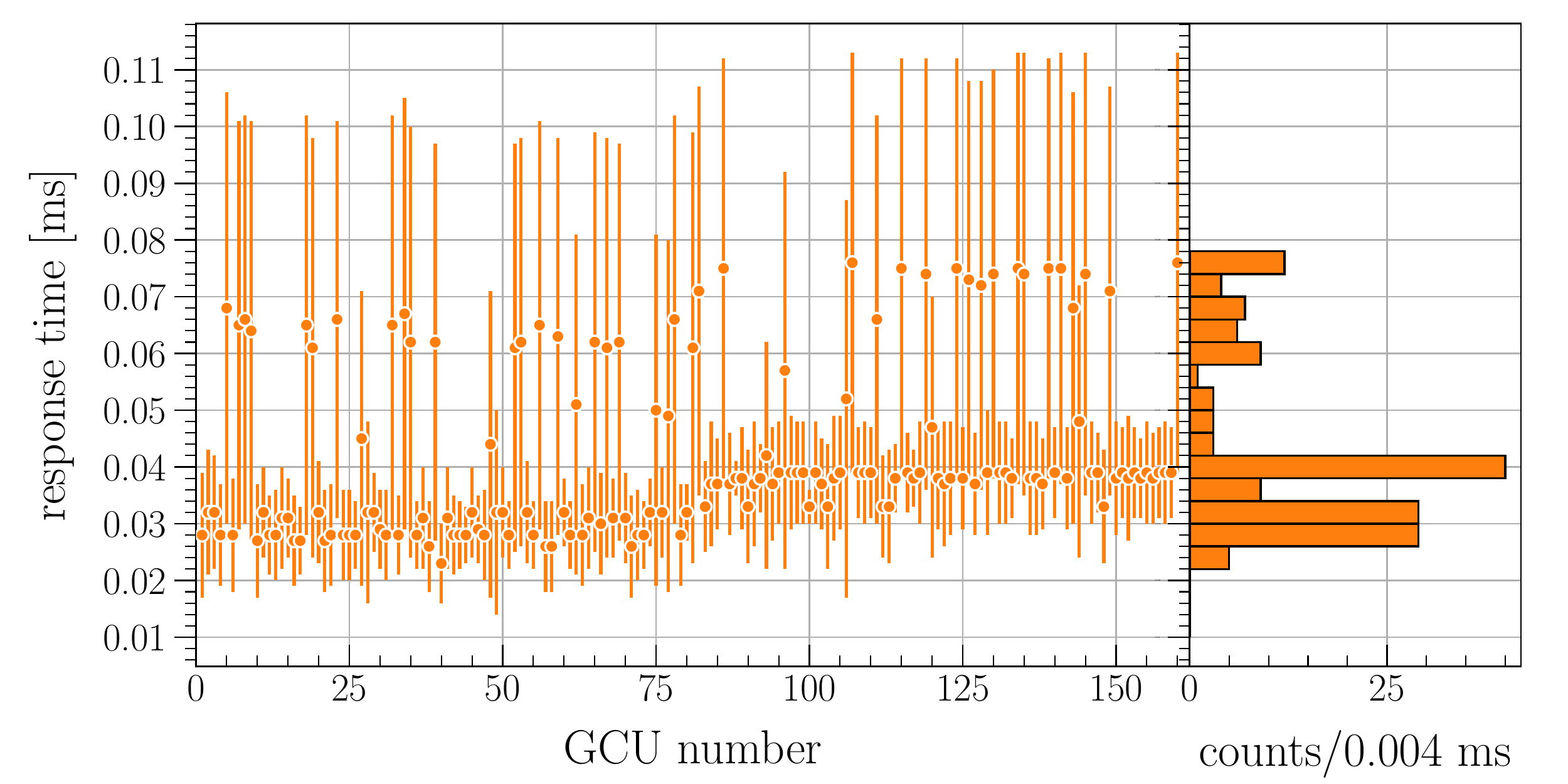}
  \caption{Ping test results for a batch of 160 GCUs. The plot on the left shows the mean response time and its standard deviation for each GCU; the plot on the right show the distribution of the response time of the batch of GCUs. A step in the response time is visible around GCU 80, pointing at differences in cable lengths and network configuration between the first 80 GCUs and the other 80 boards of the batch.}
\label{fig:ping_results}
\end{figure}

As a quick visual feedback for the shifter, the mean response time and its standard deviation were recovered and plotted versus the GCU ID number; an example with a batch of 160 GCUs is shown in Figure~\ref{fig:ping_results}.
The mean response time depends on the length of the asynchronous link cables and on the network configuration.

\subsection{Linearity test}\label{subsec:linearity}

The linearity test was meant to test the linear response of the two FADCs serving each channel and evaluate the gain factors of the two data streams in the FEC. 
The test was performed by generating PMT-like signals with the internal test pulse circuit described in Section~\ref{subsec:testpulser}.
Before this test, the channel linearity was studied with external physics sources and by reading PMT signals on a small set of boxes~\cite{bib:Cerrone, thesis:Cerrone}.

For the test, values of the test pulse amplitude were chosen to cover a wide range.
For the high-gain stream, the range starts at \SI{1}{\photoelectron} up to about \SI{160}{\photoelectron}, before the beginning of the saturation regime.
For the low-gain stream, the range starts at about \SI{90}{\photoelectron} up to the maximum possible value of the DAC, corresponding to about \SI{1200}{\photoelectron}.
The two ranges overlap, allowing us to check the cross range between the two streams.
The frequency of pulses generation and the acquisition time were set to provide more than \num{2000} waveforms for each linearity point.
Parameter settings, test pulse generation, and data acquisition were completely automatized.

Raw data were then processed and saved in \texttt{ROOT}~\cite{bib:root} files as \texttt{TTree} objects.
For each channel and each input DAC amplitude, the integrated output charge was evaluated according to eq.~\eqref{eqn:output_charge}; the evaluated values were then collected into an histogram and the mean value was taken as the output charge corresponding to the given input DAC amplitude. 
Finally, for each channel, a quadratic fit was done for both data streams to extract the gain factor of the two FEC streams, with the fit function defined as:

\begin{linenomath}
\begin{equation}
Q_{\text{out}} = c_2 \cdot Q_{\text{in}}^2 + G \cdot Q_{\text{in}} + c_0, 
\label{eqn:lin_fit}
\end{equation}
\end{linenomath}

where $Q_{\text{out}}$ and $Q_{\text{in}}$ are the output and input charge defined by eq.~\eqref{eqn:output_charge} and~\eqref{eqn:input_charge}, respectively, $G$ is the dimensionless gain of the FEC stream, $c_0$ is the intercept, and $c_2$ is the coefficient of the quadratic term.
A quadratic function was used because the response is not perfectly linear, due to the differential non-linearity (DNL) which is characteristic of ADCs and DACs; we expect the quadratic term to be subdominant with respect to the linear term.
The gain $G$ is expected to be $<1$; the reason for this design choice is that the FEC input signal is expected to reach amplitudes exceeding the typical FADC dynamic range, hence the necessity to attenuate and not amplify the signal.

\begin{figure}[hbtp] 
  \centering

	\includegraphics[width=0.8\columnwidth]{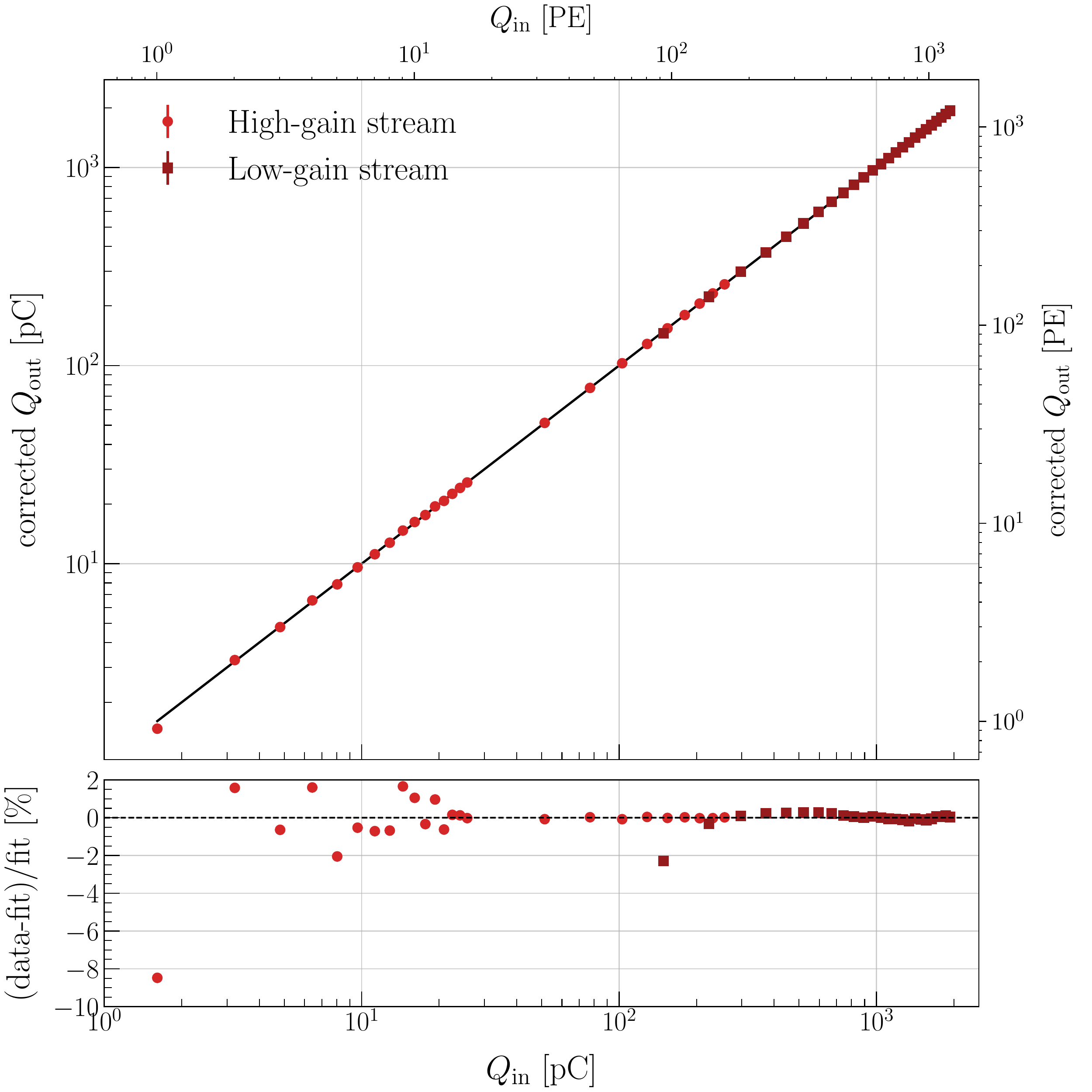} 

  \caption{Results from the linearity test for one channel of a typical GCU for the high-gain (circles, light red) and low-gain (squares, dark red) streams. Results from the quadratic fit for the high-stream are: $c_0 = (-0.01 \pm 0.02)$\,\si{\pico\coulomb}, $G=0.5856 \pm 0.0006$, and $c_2 = (\num{-4.5e-5} \pm \num{0.3e-5})$\,\si{\per\pico\coulomb}; while for the low-gain stream are: $c_0 = (0.90 \pm 0.08)$\,\si{\pico\coulomb}, $G=0.0850 \pm 0.0002$, and $c_2 = (\num{-3.3e-6} \pm \num{0.9e-6})$\,\si{\per\pico\coulomb}. The fit ranges are [1.6, 257]\,\si{\pico\coulomb} and [149, 1934]\,\si{\pico\coulomb} for the high-gain and the low-gain streams, respectively.
  The input charge is evaluated by using eq.~\eqref{eqn:input_charge}, while the output charge is first evaluated through eq.~\eqref{eqn:output_charge} and then corrected for the gain obtained from the quadratic fit. Charges are also expressed in number of PEs on the secondary axes, where $\SI{1}{\photoelectron}= \SI{1.6}{\pico\coulomb}$.}
\label{fig:linearity_both}
\end{figure}

Figure~\ref{fig:linearity_both} shows the results of the linearity test for one channel of a typical GCU for the high-gain stream (circles, light red) and the low-gain stream (squares, dark red), respectively. 
The corrected output charge shown in the plot was first evaluated through eq.~\eqref{eqn:output_charge} and then corrected with the gain obtained from the quadratic fit; as it can be seen, after the gain correction the two data streams lie on the same line.
In the figure, input and output charges are expressed in picocoulomb on the primary axes and in terms of number of photo-electrons (PE) on the secondary axes, with $\SI{1}{\photoelectron} = q_e \cdot G_{\mathrm{PMT}} = \SI{1.6}{\pico\coulomb}$, where $q_e$ is the electron charge, and $G_{\mathrm{PMT}} = \num{e7}$ is the assumed nominal PMT gain of the 20-inch PMTs in JUNO~\cite{bib:juno:phys-det,bib:pmt_testing}.

During the analysis, we also checked for the saturation amplitude of the high-gain stream, while for the low-gain stream we could not reach saturation with the internal test pulse generator. 
In the high-gain configuration, channels saturate for an input signal of about 16500 DAC counts, corresponding to an input charge of about \SI{450}{\pico\coulomb} $\simeq$ \SI{280}{\photoelectron}. 
Data points above the saturation threshold are not used in the linear fit and are not shown in Figure~\ref{fig:linearity_both}.

\subsection{Stability test}\label{subsec:stability}

The stability test consists in firing the internal test pulse generator with a fixed amplitude over a time period lasting several hours, and to check that the waveform properties listed below do not change. 
The input amplitude was set to 12000 DAC counts for the high-gain stream and to 45000 DAC counts for the low-gain stream.
The frequency of the test pulses was set to \SI{1}{\hertz}, while the data acquisition time was determined by the available time during the shift.

The waveform monitored parameters are: baseline, noise, minimum value of the waveform, and minimum position in the readout window.
The baseline and noise are obtained as described in Section~\ref{subsec:waveform}.
These quantities were obtained by processing raw data and saved in \texttt{ROOT} files as \texttt{TTree} objects.

\begin{figure}[htbp] 
  \centering
  \includegraphics[width=1\columnwidth]{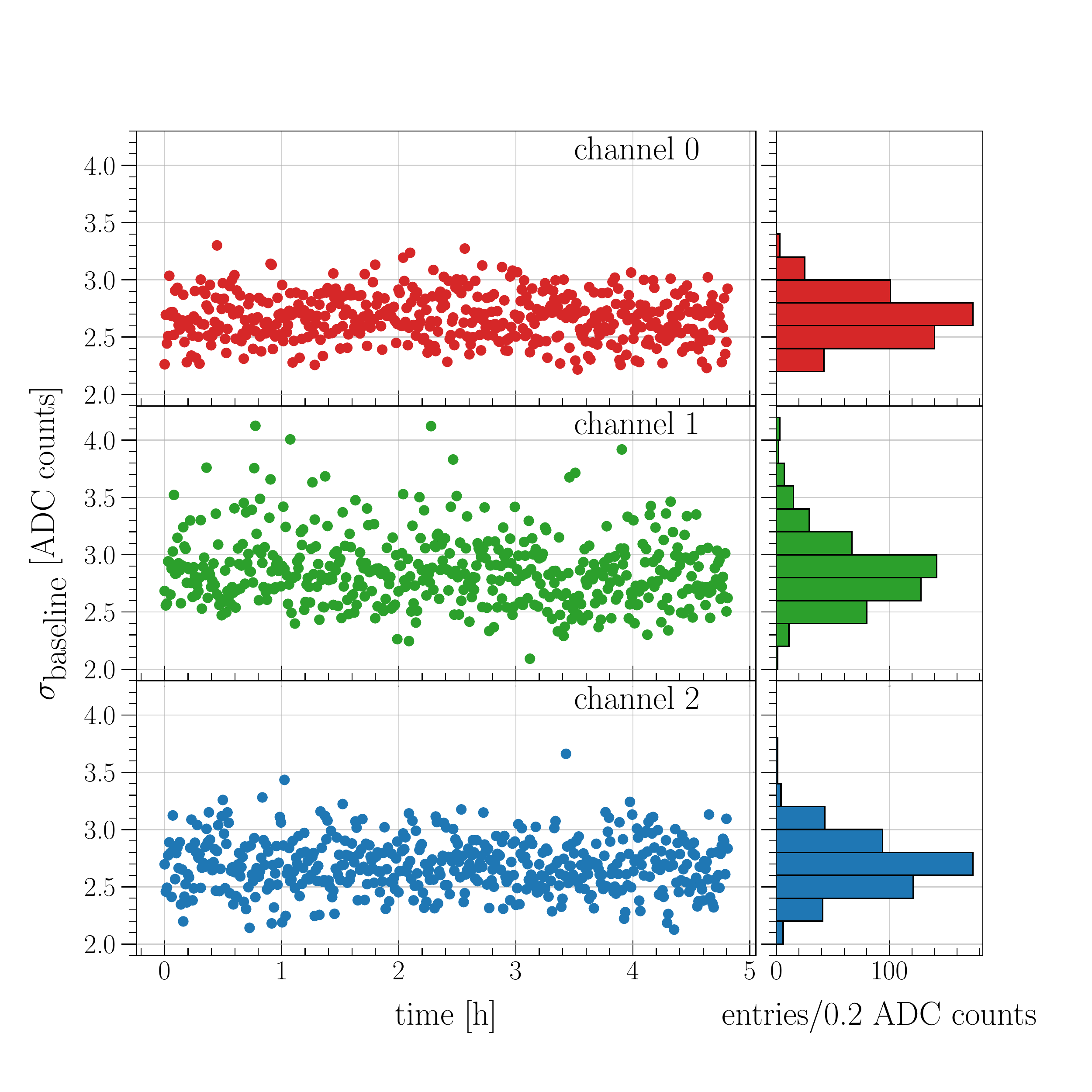}
  \caption{Evolution of the noise over a 5-hour stability run for the three channels of a typical GCU. The left plots show the noise evaluated on single waveforms as a function of time; the right plots show the distribution of the noise values. For all three channels, noise is within the acceptance interval.}
\label{fig:stability_bsl_sigma}
\end{figure}

As an example, Figure~\ref{fig:stability_bsl_sigma} shows the results of the stability test for the noise of a typical GCU. 
The value of the noise as a function of time is shown for the three channels in three different panels; distributions of the values are shown as well.
The accepted noise level is between 2 and 4.5 ADC counts, corresponding to about \SI{0.03}{\photoelectron} and \SI{0.08}{\photoelectron} respectively, and, as it can be seen in Figure~\ref{fig:stability_bsl_sigma}, the evaluated values lie within these limits.

\subsection{Slow control monitoring}\label{subsec:slowcontrol}

The slow control monitoring is meant to read several internal parameters and sensors installed on the GCU and to monitor the overall status of the board. 
All sensors were read through the IPbus protocol~\cite{bib:ipbus} in parallel to the DAQ and over the same transport layer.

For each GCU, the following parameters were read during the slow control monitoring: the temperature of the FPGA, the temperature and the high voltage value of each HVU, and several FPGA internal reference voltages~\cite{bib:sensors}.

\begin{figure}[hbtp] 
  \centering
  \includegraphics[width=1\columnwidth]{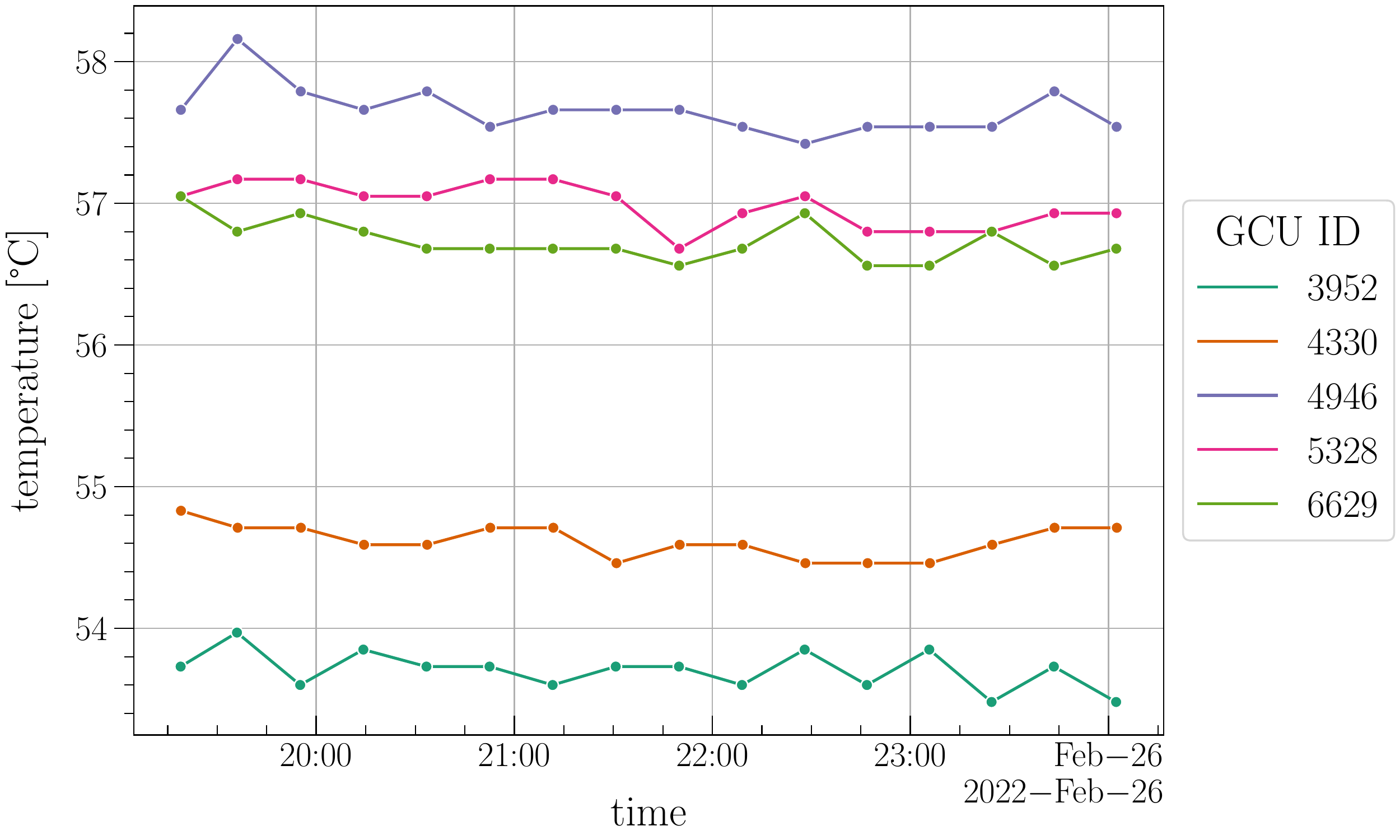}
  \caption{The figure shows the evolution of the FPGA temperature for 5 GCUs, recorded from the slow control monitoring.  The different temperature values are due to the different positions of the GCUs in the testing room in the dedicated facility at Kunshan.}
\label{fig:slow_control}
\end{figure}

As an example, Figure~\ref{fig:slow_control} shows a plot of the evolution of the FPGA temperature for five GCUs. 
For all GCUs, the FPGA temperature is stable over time. 
The difference in the absolute values is due to the different positions of the GCUs on the racks in the testing room (see Section~\ref{sec:kunshan}). 
The testing room was equipped with an air conditioning system with a constant temperature of about \SI{26}{\celsius}.

\subsection{Storing of test results into a database}\label{subsec:database}

The information on the configuration and parameters used for the tests, together with the results of the tests, are saved in a MySQL database which is available on the local server at the Kunshan site.
Storing these kinds of information is important to have an history of the performances of each GCU, and to compare the results during mass production with the tests foreseen for the upcoming installation and commissioning phases.

\begin{figure}[hbtp] 
  \centering
  \includegraphics[width=0.8\columnwidth]{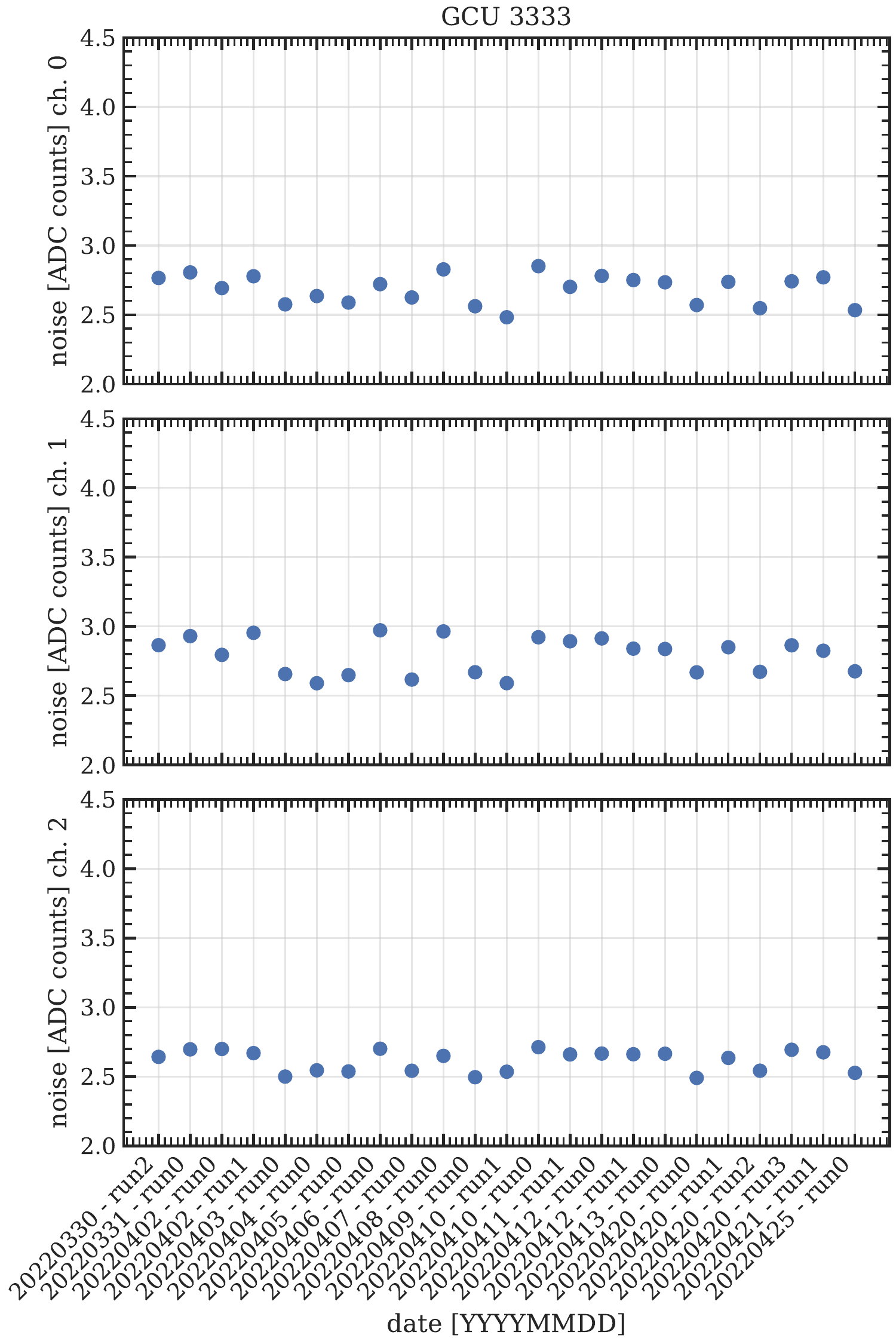}
  \caption{Results of the stability test from several runs for GCU 3333 are shown. The three panels show the values of the noise for channel 0 (top), channel 1 (middle) and channel 2 (bottom). The noise for all channels is stable and within the acceptance range over a period of more than 25 days.
}
\label{fig:database}
\end{figure}

Figure~\ref{fig:database} is obtained by accessing the local database and shows the value of the noise from the stability test for several days and runs for GCU 3333; each panel shows results for one of the three GCU channels. 
The runs shown in the figure span a time period of more than 25 days, during which the noise is stable and within the acceptance range.

\begin{figure}[hbtp] 
  \centering
		\subfloat[][\emph{Gain distribution for the high-gain stream.}]
		   {\includegraphics[width=0.7\columnwidth]{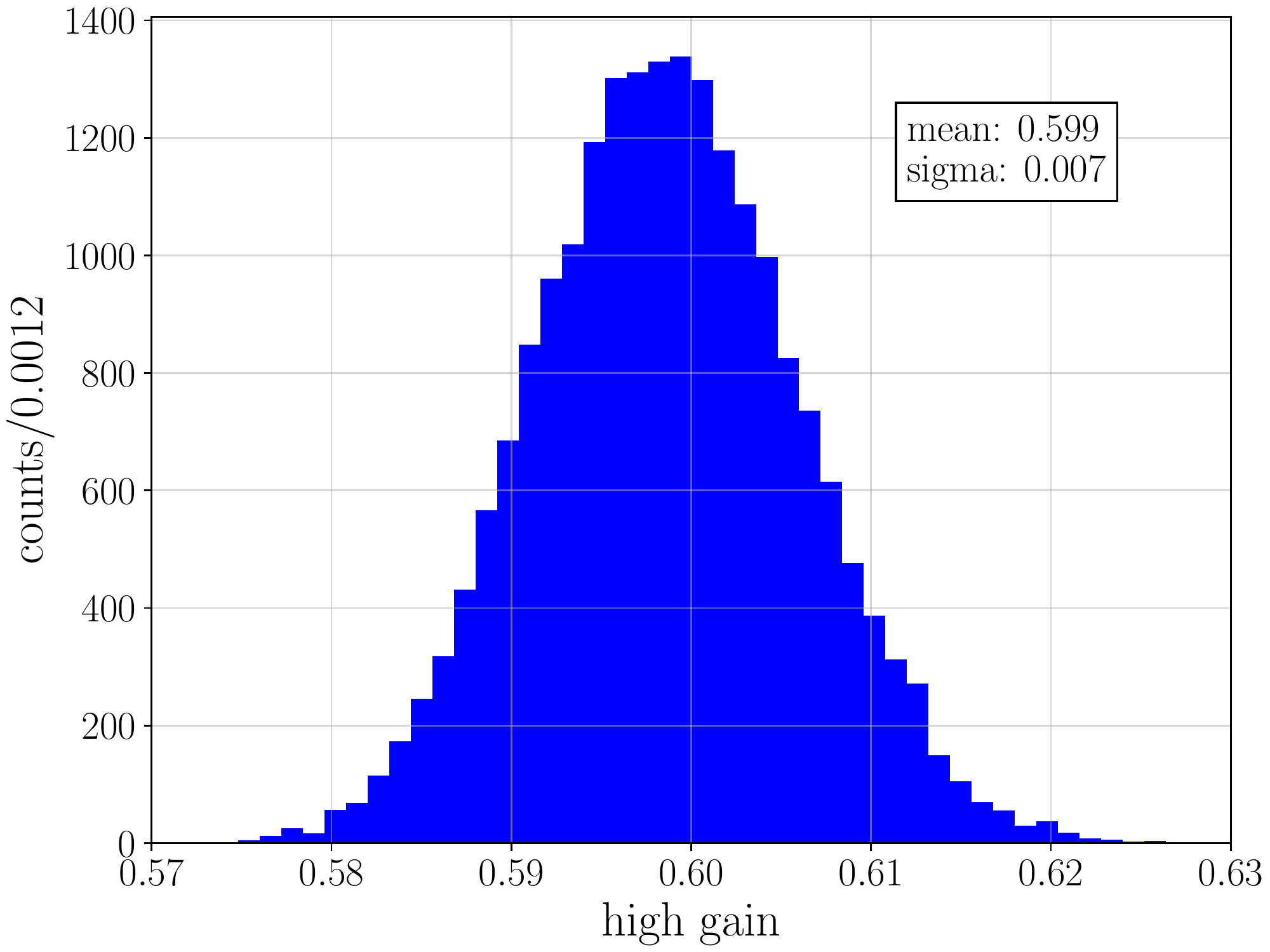}
		   \label{subfig:highgain}} \\
		\subfloat[][\emph{Gain distribution for the low-gain stream.}]
		   {\includegraphics[width=0.7\columnwidth]{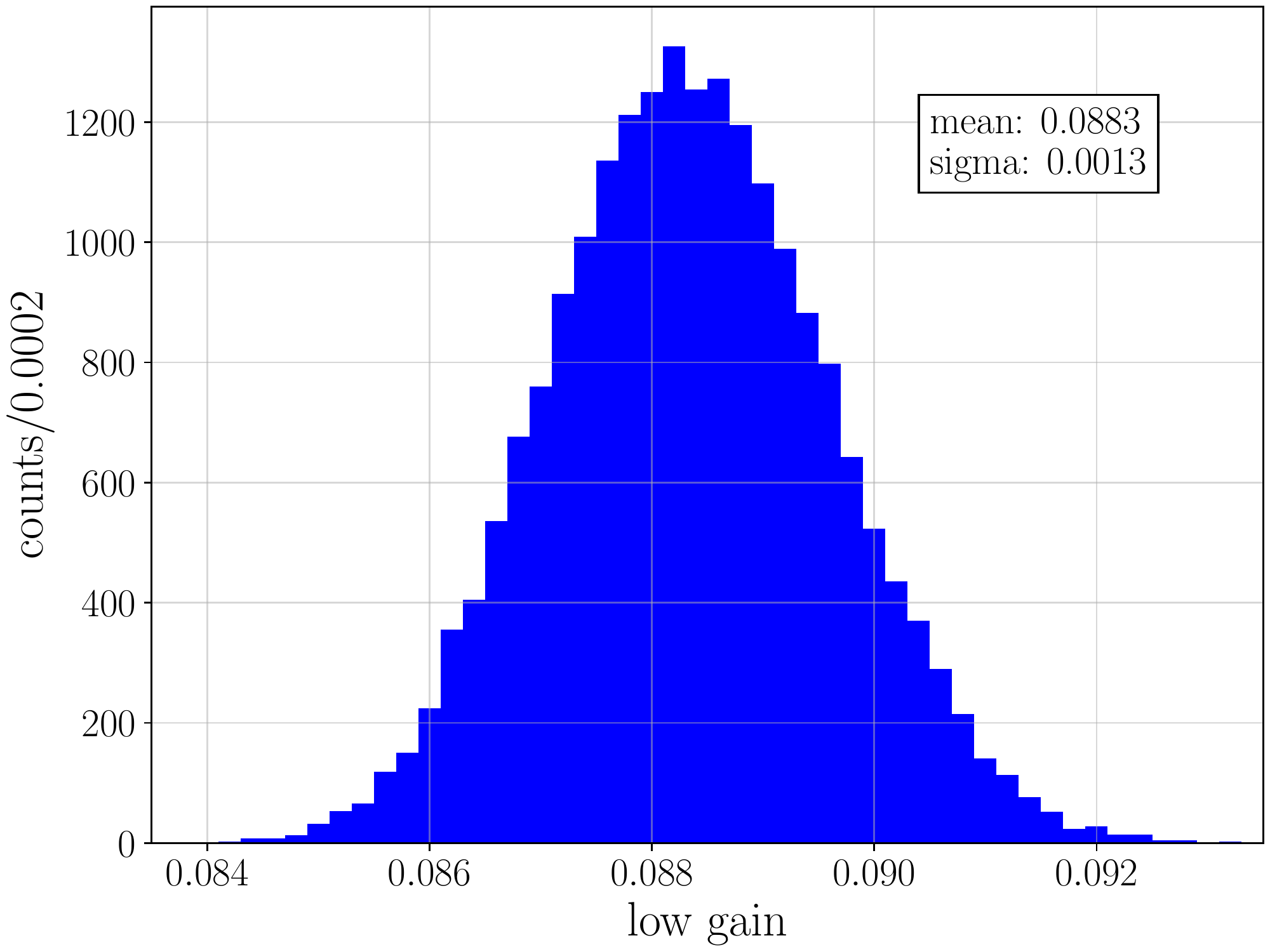}
		   \label{subfig:lowgain}} \\  

  \caption{Distributions of the gain obtained from the linearity test for (a) the high-gain stream and (b) the low-gain stream. The distribution of the high gain has mean and standard deviation equal to 0.599 and 0.007, respectively; the distribution of the low gain has mean and standard deviation equal to 0.0883 and 0.0013, respectively.}
\label{fig:gain_distribution}
\end{figure}

Figures~\ref{subfig:highgain} and~\ref{subfig:lowgain} show the distributions of the high-gain and the low-gain values, respectively, obtained in the linearity test by using eq.~\eqref{eqn:lin_fit}.
The distribution for the high-gain stream has a mean of 0.599 and a standard deviation of 0.007, while the distribution for the low-gain stream has a mean of 0.0883 and a standard deviation of 0.0013.

\section{Conclusion}\label{sec:conclusion}

\begin{table}[ht]
\caption{Acceptance range for the baseline, noise, high gain, and low gain, used as acceptance criteria for the evaluation of the performance of each GCU.}
\centering
\begin{tabular}{ c | c  }
 Parameter & Acceptance range  \\ 
 \hline
 baseline & 11000 - 12000 ADC counts \\  
 noise & 2 - 4.5 ADC counts \\
 high gain & 0.5 - 0.65 \\
 low gain & 0.05 - 0.095
\end{tabular}
\label{tab:acceptance}
\end{table}

A test protocol was developed to evaluate the performance of the 20-inch PMT readout electronics for the JUNO experiment during mass production.
A total of 6950 devices were tested in about ten months. 
Only eight GCUs were discarded on the basis of the tests presented in this work and the criteria shown in Table~\ref{tab:acceptance}.
Other 56 GCUs were discarded due to issues arisen during the assembling procedure.
In total, 6886 GCUs were accepted, while only 64 were rejected, providing a final acceptance yield of \SI{99.1}{\percent}.
Out of the 6886 accepted cards, 6681 will be used in the CD and Water Pool veto system, 25 will be used by OSIRIS~\cite{OSIRIS}, while the remaining 180 will be kept as backup.
The test protocol described in this paper will be used as a reference for the upcoming tests during the installation and commissioning phases, where a few adjustments are needed given the different environmental conditions and setup.

\section*{Acknowledgements}
Part of this work has been supported by the Italian-Chinese
collaborative research program jointly funded by the Italian Ministry of
Foreign Affairs and International Cooperation (MAECI) and the National
Natural Science Foundation of China (NSFC).

\bibliographystyle{elsarticle-num}
\bibliography{main}

\begin{thebibliography}{10}
\expandafter\ifx\csname url\endcsname\relax
  \def\url#1{\texttt{#1}}\fi
\expandafter\ifx\csname urlprefix\endcsname\relax\def\urlprefix{URL }\fi
\expandafter\ifx\csname href\endcsname\relax
  \def\href#1#2{#2} \def\path#1{#1}\fi

\bibitem{bib:juno:phys-det}
A.~Abusleme, et~al., {JUNO physics and detector}, Prog. Part. Nucl. Phys. 123
  (2022) 103927.
\newblock \href {http://arxiv.org/abs/2104.02565} {\path{arXiv:2104.02565}},
  \href {http://dx.doi.org/10.1016/j.ppnp.2021.103927}
  {\path{doi:10.1016/j.ppnp.2021.103927}}.

\bibitem{bib:juno:yb}
F.~An, et~al., {Neutrino Physics with JUNO}, J. Phys. G 43~(3) (2016) 030401.
\newblock \href {http://arxiv.org/abs/1507.05613} {\path{arXiv:1507.05613}},
  \href {http://dx.doi.org/10.1088/0954-3899/43/3/030401}
  {\path{doi:10.1088/0954-3899/43/3/030401}}.

\bibitem{bib:precision}
{JUNO Collaboration}, A.~Abusleme, et~al., {Sub-percent Precision Measurement
  of Neutrino Oscillation Parameters with JUNO}, submitted to Chinese Physics C
  (2022).
\newblock \href {http://dx.doi.org/10.48550/ARXIV.2204.13249}
  {\path{doi:10.48550/ARXIV.2204.13249}}.

\bibitem{bib:juno_pmt}
L.-J. Wen, M.~He, Y.-F. Wang, J.~Cao, S.-L. Liu, Y.-K. Heng, Z.-H. Qin, {A
  quantitative approach to select PMTs for large detectors}, Nucl. Instrum.
  Meth. A 947 (2019) 162766.
\newblock \href {http://arxiv.org/abs/1903.12595} {\path{arXiv:1903.12595}},
  \href {http://dx.doi.org/10.1016/j.nima.2019.162766}
  {\path{doi:10.1016/j.nima.2019.162766}}.

\bibitem{bib:pmt_testing}
A.~Abusleme, et~al., {Mass Testing and Characterization of 20-inch PMTs for
  JUNO}, submitted to European Physical Journal C (2022).
\newblock \href {http://dx.doi.org/10.48550/ARXIV.2205.08629}
  {\path{doi:10.48550/ARXIV.2205.08629}}.

\bibitem{CAO2021165347}
C.~Cao, et~al.,
  \href{https://www.sciencedirect.com/science/article/pii/S0168900221003314}{Mass
  production and characterization of 3-inch pmts for the juno experiment},
  Nuclear Instruments and Methods in Physics Research Section A: Accelerators,
  Spectrometers, Detectors and Associated Equipment 1005 (2021) 165347.
\newblock \href {http://dx.doi.org/https://doi.org/10.1016/j.nima.2021.165347}
  {\path{doi:https://doi.org/10.1016/j.nima.2021.165347}}.
\newline\urlprefix\url{https://www.sciencedirect.com/science/article/pii/S0168900221003314}

\bibitem{bib:elec:bx}
M.~Bellato, et~al., {Embedded readout electronics R\&D for the large PMTs in
  the JUNO experiment}, Nucl. Instrum. Meth. A 985 (2021) 164600.
\newblock \href {http://arxiv.org/abs/2003.08339} {\path{arXiv:2003.08339}},
  \href {http://dx.doi.org/10.1016/j.nima.2020.164600}
  {\path{doi:10.1016/j.nima.2020.164600}}.

\bibitem{bib:electronics}
{The JUNO collaboration}, {The JUNO experiment large-PMT read-out and trigger
  electronics}, paper in preparation.

\bibitem{thesis:Coppi}
A.~Coppi, \href{http://hdl.handle.net/20.500.12608/1067}{{Design and
  development of Data Quality Monitoring protocols for the integration tests of
  the JUNO large PMT electronics}}, {Bachelor's thesis, University of Padua,
  Italy} (2021).
\newline\urlprefix\url{http://hdl.handle.net/20.500.12608/1067}

\bibitem{thesis:Marini}
F.~Marini, {Development and Testing of the large PMTs Front-End Electronics for
  the JUNO Experiment}, Ph.D. thesis, University of Padua, Italy (2021).

\bibitem{Liu:2022nhe}
C.~Liu, et~al., {Check on the features of potted 20-inch PMTs with 1F3
  electronics prototype at Pan-Asia}, submitted to Radiation Detection
  Technology and Methods (8 2022).
\newblock \href {http://arxiv.org/abs/2208.08264} {\path{arXiv:2208.08264}}.

\bibitem{thesis:Triozzi}
R.~Triozzi, \href{http://hdl.handle.net/20.500.12608/22257}{{Supernova neutrino
  detection in JUNO, a large liquid scintillator neutrino detector}},
  {Bachelor's thesis, University of Padua, Italy} (2021).
\newline\urlprefix\url{http://hdl.handle.net/20.500.12608/22257}

\bibitem{Gong:2015yvq}
G.~Gong, H.~Gong, H.~Li, T.~Xue, {The Global Trigger with Online Vertex Fitting
  for Low Energy Neutrino Research}, in: {15th International Conference on
  Accelerator and Large Experimental Physics Control Systems}, 2015, p.
  THHB2O03.
\newblock \href {http://dx.doi.org/10.18429/JACoW-ICALEPCS2015-THHB2O03}
  {\path{doi:10.18429/JACoW-ICALEPCS2015-THHB2O03}}.

\bibitem{bib:ipbus}
C.~Ghabrous~Larrea, K.~Harder, D.~Newbold, D.~Sankey, A.~Rose, A.~Thea,
  T.~Williams, {IPbus: a flexible Ethernet-based control system for xTCA
  hardware}, JINST 10~(02) (2015) C02019.
\newblock \href {http://dx.doi.org/10.1088/1748-0221/10/02/C02019}
  {\path{doi:10.1088/1748-0221/10/02/C02019}}.

\bibitem{bib:Triozzi}
R.~Triozzi, et~al., {Implementation and performance of IPbus in JUNO Large-PMT
  data acquisition stream}, paper in preparation.

\bibitem{bib:Cerrone}
V.~Cerrone, et~al., {Validation and integration tests of the JUNO 20-inch PMTs
  readout electronics}, paper in preparation.

\bibitem{thesis:Cerrone}
V.~Cerrone, \href{http://hdl.handle.net/20.500.12608/936}{{Characterization of
  the final read-out electronics for the large PMTs of the JUNO experiment}},
  {Bachelor's thesis, University of Padua, Italy} (2021).
\newline\urlprefix\url{http://hdl.handle.net/20.500.12608/936}

\bibitem{bib:root}
R.~Brun, F.~Rademakers, {ROOT - An Object Oriented Data Analysis Framework},
  Nucl. Inst. \& Meth. in Phys. Res. A~(389) (1996) 81--86, proceedings
  AIHENP'96 Workshop, Lausanne, Sep. 1996.

\bibitem{bib:sensors}
Xilinx,
  \href{https://docs.xilinx.com/v/u/en-US/ds182_Kintex_7_Data_Sheet.pdf}{Kintex‐7
  FPGAs Data Sheet: DC and AC Switching Characteristics} (2021).
\newline\urlprefix\url{https://docs.xilinx.com/v/u/en-US/ds182_Kintex_7_Data_Sheet.pdf}

\bibitem{OSIRIS}
A.~Abusleme, et~al., The design and sensitivity of juno’s scintillator
  radiopurity pre-detector osiris, Eur. Phys. J. C 81 (2021) 973.
\newblock \href
  {http://dx.doi.org/https://doi.org/10.1140/epjc/s10052-021-09544-4}
  {\path{doi:https://doi.org/10.1140/epjc/s10052-021-09544-4}}.

\end{thebibliography}

\end{document}